\documentclass[twocolumn,showpacs,preprintnumbers,amsmath,amssymb]{revtex4}
\usepackage{graphicx}% Include figure files
\usepackage{dcolumn}% Align table columns on decimal point
\usepackage{bm}% bold math

\begin{document}

\preprint{}

% \title{From Localized Moments to the Heavy Fermi Liquid in the Kondo Lattice Model}% Force line breaks with \\
\title{Evolution of a Large Fermi Surface in the Kondo Lattice}% Force line breaks with \\
%.\title{Evolution of Fermi Liquid with the Large Fermi Surface in the Kondo Lattice}% Force line breaks with \\

\author{Junya Otsuki$^1$}
%  \altaffiliation[Also at ]{Physics Department, XYZ University.}%Lines break automatically or can be forced with \\
\author{Hiroaki Kusunose$^2$}
\author{Yoshio Kuramoto$^1$}
\affiliation{
$^1$Department of Physics, Tohoku University, Sendai 980-8578, Japan\\
$^2$Department of Physics, Ehime University, Matsuyama 790-8577, Japan
}%

\date{\today}% It is always \today, today,

\begin{abstract}
Single-particle spectrum of the Kondo lattice model is 
derived with use of the continuous-time quantum Monte Carlo method, 
combined with the dynamical mean-field theory.
%Evolution of either a Fermi-liquid state 
%or a magnetically ordered state is demonstrated from the high-temperature local-moment regime.
%.
Crossover behavior is traced quantitatively either to a heavy Fermi-liquid state or to a magnetically ordered state 
from the local-moment state at high temperatures.
The momentum distribution in the low-temperature limit
acquires a discontinuity at the location
that involves the local-spin degrees of freedom.
%.
Even without the charge degrees of freedom for local electrons,
the excitation spectra exhibit 
hybridized bands similar to those in the Anderson lattice.
Temperature dependence in the 
zero-energy component of the self-energy is crucial in forming the Fermi-liquid state with the large Fermi surface.
\end{abstract}

\maketitle

%An interesting situation 
A long-standing problem in condensed matter physics 
%often occurs when two competing effects meet together.
is how to resolve the dichotomy between localized and 
itinerant characters of electrons in solids.
For many metals with 
%$d$ or 
$f$ 
%valence 
electrons, and also some with $d$ electrons,
the strong Coulomb repulsion between localized electrons
suppresses charge fluctuations at each site, 
%for them, emphasizing 
% makes them localized with
leaving only a spin and/or an orbital degrees of freedom.
These localized degrees of freedom
%, which
interact with conduction electrons,
which are delocalized over the entire crystal.
The itinerant and localized electrons affect each other, and realize rich physical phenomena such as heavy electrons.

The simplest 
%fundamental 
model to describe the situation is the Kondo lattice model (KLM) given by 
\begin{equation}
H = \sum_{{\bm k} \sigma} \epsilon_{{\bm k}} c_{{\bm k} \sigma}^{\dag} c_{{\bm k} \sigma}
+ J \sum_i {\bm S}_i \cdot {\bm \sigma}_i.
\end{equation}
Here ${\bm S}_i$ represents the localized spin of the valence electron at the $i$ site, and ${\bm \sigma}_i=\sum_{\sigma \sigma'} c^{\dag}_{i\sigma} {\bm\sigma}_{\sigma \sigma'} c_{i\sigma'}$ is the spin operator of the itinerant conduction electron.
%Despite the apparently simple model, it gives an account of diverse phenomena due to the localized nature.
%
The antiferromagnetic exchange interaction $J>0$ 
under realistic condition is 
%usually weak 
much smaller than the bandwidth, and each localized spin acts as a weak scatterer for the conduction electrons at high temperatures.
As temperature decreases, interactions 
between the localized spins become significant,  
which are mediated by conduction electrons,
%via the Ruderman-Kittel-Kasuya-Yosida (RKKY) interaction, which favors 
and a magnetic long-range order may be realized.
%For moderately large $J$, on the other hand,
On the other hand, 
%if $J$ is large enough,
the localized spins also tend to be quenched by conduction electrons, which is called the Kondo effect.
If the system remains paramagnetic,
coherent quasi-particles may emerge by
the collective Kondo effect.
Although the overall 
picture~\cite{Doniach} mentioned above is widely 
accepted, the rich crossover phenomena 
due to emergent quasi-particles
remain highly nontrivial~\cite{Kuramoto}.
% evolution of the quasi-particle states
%the nature of the low-temperature states, 
%including the size of the Fermi surface, 
%with the competition between the 
%Ruderman-Kittel-Kasuya-Yosida (RKKY) interaction and the Kondo effect,
%The formation process of quasi-particles is relevant
%to observable quantitities such as the resistivity,
%probed 
% the de Haas-van Alphen effect, 
%for example~\cite{Stewart, Kuramoto}.
%%in strongly correlated electron systems~\cite{Stewart, Degiorgi, Allen}.

The KLM has no apparent counterpart of non-interacting system, since the perturbation series with respect to $J$ is essentially singular at $J=0$.
Accordingly, we have no clear starting point to study the nature of 
%Fermi-liquid 
the paramagnetic ground state in the KLM.
On the other hand, 
an ordinary Fermi-liquid argument is applicable to 
 the 
% periodic Anderson model (Anderson lattice),
Anderson lattice,
since the adiabatically continued non-interacting 
ground state is evident~\cite{Yamada-Anderson-lattice}.
%In particular, explicit investigations have been performed
%using the dynamical mean-field theory (DMFT), in which the local dynamics can beis rigorously taken into account~\cite{Georges}. An important 
A clue for characterizing the KLM is the Luttinger's theorem
%, which is the fundamental result on the Fermi liquid
which relates the volume of the Fermi surface with the number of electrons~\cite{Luttinger}.
A version of proof~\cite{Oshikawa} of this theorem 
states
that the volume of the Fermi-surface is the same as that in
%contains the localized spins as similar to 
the Anderson lattice (the so-called ``large Fermi surface") with unit occupation of each local state.
The proof is 
%{\it even} 
valid 
%{\it 
provided that the ground state of the KLM is a Fermi liquid.
%}. %The topological 
%.
However, it is not trivial whether this condition is satisfied.
In fact, another possibility of a ``small" Fermi surface has also been proposed~\cite{coleman,senthil,burdin} with inclusion of explicit coupling between local spins.

Because of the difficulty in including higher-order effects of $J$,
the KLM has been studied either by variants of mean-field theory~\cite{lacroix,Shiba-Fazekas,burdin},
or in one dimensional systems~\cite{Moukouri, Tsunetsugu97, Shibata99}.
%~\cite{Ueda-FS, Moukouri, Tsunetsugu97, Shibata99}.
%.
In the mean-field theory, the collective Kondo effect appears as a phase transition,
and 
%..
it is difficult to reproduce the Kondo crossover at finite temperatures.
As a particular feature in one dimension, on the other hand,
the low-energy excitations behave as in a Tomonaga-Luttinger liquid without discontinuity in the momentum distribution.
Furthermore, a magnetic long-range order, which competes with the paramagnetic state, is 
%completely 
strongly suppressed.
Thus it is highly desirable to obtain reliable information in
higher dimensions, especially the temperature-dependent evolution of the collective Kondo effect.

In this Letter, 
we report results which become exact in the limit of infinite dimensions.  
First, evolution of the Landau quasi-particles is 
%demonstrated by starting
%..
traced quantitatively from the local-moment regime at high temperatures. 
Secondly, magnetic phase diagram of the KLM is derived.
%Single-particle excitation spectrum and the momentum distribution are derived in the framework of the 
The periodic lattice effect is dealt with by the 
dynamical mean-field theory (DMFT)~\cite{Georges}.
To solve the effective impurity problem in the DMFT,
we use the continuous-time quantum Monte Carlo (CT-QMC) 
algorithm~\cite{Rubtsov,Werner},
adapted to the Kondo model~\cite{Otsuki-CTQMC}.
Since no approximation such as discretization is involved in the CT-QMC algorithm, and since the simulation does not encounter the minus sign problem, the effective impurity problem can be solved 
%exactly within error bars.
with desired accuracy.

{\bf Model and method.} ---
Let us start with 
%In the KLM, all the information of the single-particle excitations is contained in
the conduction-electron Green function in the KLM:
\begin{equation}
G_{\rm c}({\bm k}, i\epsilon_n) = [i\epsilon_n -\epsilon_{\bm k} + \mu - 
%\Sigma_{\rm c}({\bm k}, i\epsilon_n)]^{-1},
\Sigma_{\rm c}(i\epsilon_n)]^{-1},
\label{green}
\end{equation}
where $\epsilon_n=(2n+1)\pi T$ is the fermionic Matsubara frequency. % for fermions.
The self-energy $\Sigma_{\rm c}(i\epsilon_n)$ takes 
%full 
account of non-perturbative contributions from 
%expresses influences of
the localized spins, 
but the momentum dependence has been neglected according to the DMFT.
%In the DMFT, the ${\bm k}$-dependence of $\Sigma_{\rm c}({\bm k}, i\epsilon_n)$ is negligible.
%completely neglected, which becomes exact in infinite dimension.
The KLM is then mapped to the impurity Kondo model in the effective medium, which is 
%expressed 
characterized by the cavity Green function, 
$\mathcal{G}_{\rm c}^0(i\epsilon_n)$.
%represents a site-diagonal component of the `bare' Green function.
The impurity self-energy $\Sigma_{\rm c}(i\epsilon_n)$ 
%, that is to be identical with the approximate lattice self-energy, 
should be identical with that in Eq.~(\ref{green}),
%the approximate lattice self-energy, 
and is given in terms of the impurity $t$-matrix $t(i\epsilon_n)$ as $\Sigma_{\rm c}(i\epsilon_n)^{-1} = t(i\epsilon_n)^{-1} + \mathcal{G}_{\rm c}^0(i\epsilon_n)$. 
%.
Here $\mathcal{G}_{\rm c}^0(i\epsilon_n)$ is determined from the local Green function 
$\bar{G}_{\rm c}(i\epsilon_n)$ of the lattice system 
via $\mathcal{G}_{\rm c}^0(i\epsilon_n)^{-1}= \bar{G}_{\rm c}(i\epsilon_n)^{-1}+\Sigma_{\rm c}(i\epsilon_n)$.
%We search for the solution of the DMFT equations in a self-consistent manner.

% We solve the effective impurity model by means of the CT-QMC for the Kondo impurity~\cite{Otsuki-CTQMC}.
%recently developed continuous-time quantum Monte Carlo (QMC) method~\cite{Rubtsov,Werner,Otsuki-CTQMC}.
%Since the simulation does not encounter negative weight configurations, we can evaluate accurate single-particle spectra down to low enough temperatures.
We adopt a nearest-neighbor tight-binding model in the infinite-dimensional hyper-cubic lattice~\cite{Muller-Hartmann}, which is characterized by the Gaussian density of states: $\rho_{\rm c}(\omega)=D^{-1} \sqrt{2/\pi} \exp(-2\omega^2/D^2)$. 
We take $D=1$ as the unit of energy, and fix the conduction-electron density per site as $n_{\rm c}=0.9$ throughout this paper.
%.
The latter choice favors antiferromagnetism by the nearly nesting condition.
We use $10^7$ QMC samples at most with 10--100 intervals.
The DMFT self-consistent equations 
%usually converged 
converge typically within 10 iterations.

\begin{figure}[tb]
\includegraphics[width=8.5cm]{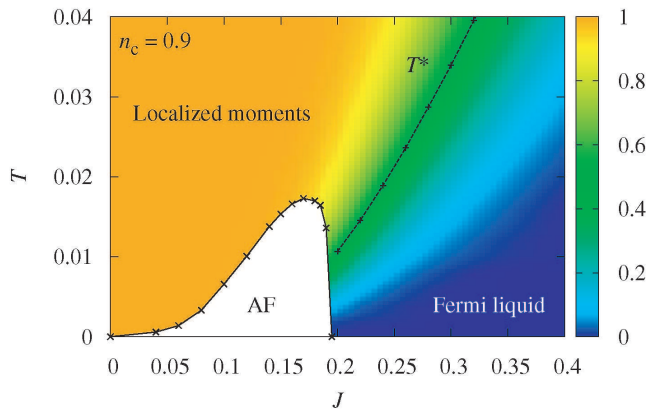}
\caption{The phase diagram of the KLM at $n_{\rm c}=0.9$. 
The intensity map represents $\xi$ defined by Eq.~(\ref{xi}),
%the measure of ``localness" of the electronic states, $\xi$, 
and the coherent energy scale $T^*$ is defined by $\xi=0.5$.}
\label{fig:phase_diagram}
\end{figure}

{\bf Phase diagram.} ---
We first show the $J$-$T$ phase diagram in Fig.~\ref{fig:phase_diagram}.
The antiferromagnetic transition temperature $T_{\rm AF}$ is determined by the divergence of the 
staggered magnetic susceptibility, which can be computed 
%in the standard procedure of 
in terms of the Bethe-Salpeter equation with numerically derived local vertices~\cite{Georges,Otsuki-KLM}.
%..
Here we have neglected possible incommensurate order for simplicity, since our main interest in this paper lies in the paramagnetic phase.
The overall structure of the phase diagram is understood by competition between the Kondo effect and the RKKY interaction~\cite{Doniach}.
%.
As $J$ increases ($T$ decreases) in the paramagnetic phase, 
%.the electronic state gradually changes 
the localized moments gradually disappear, and 
the Fermi-liquid behavior dominates below the coherent energy scale $T^*$.
The intensity map in Fig.~\ref{fig:phase_diagram} represents 
%a measure of ``localness" of the electronic states 
the degree of itinerancy 
in terms of the parameter $\xi$,
whose definition is given later in Eq.~(\ref{xi}).
%.
% , about $0.1$ in $J$ or $T$ axes
We will show that the Fermi liquid is indeed realized for $J>J_{\rm c} \simeq 0.2$ at low temperatures.
The crossover region in temperature 
becomes narrower near the quantum critical point, $J_{\rm c}$.
%Note that it is a nontrivial issue whether the low-temperature strong-coupling regime belongs to the Fermi liquid.
%, since the multiple scattering off the localized spins becomes important in this regime, and an adiabatically continued non-interacting system done not exist for the KLM.

%
\begin{figure}[tb]
\includegraphics[width=8.5cm]{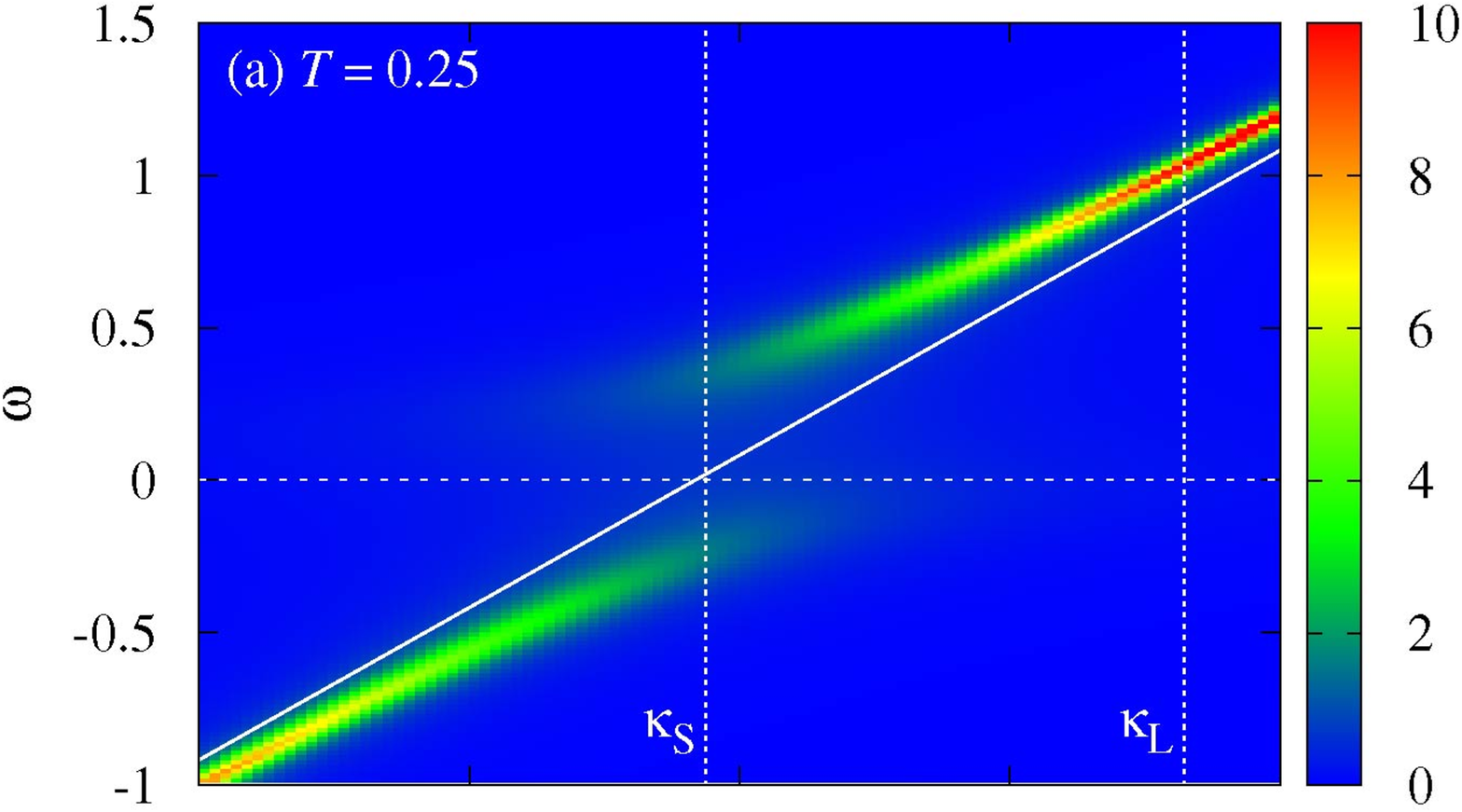}
\includegraphics[width=8.5cm]{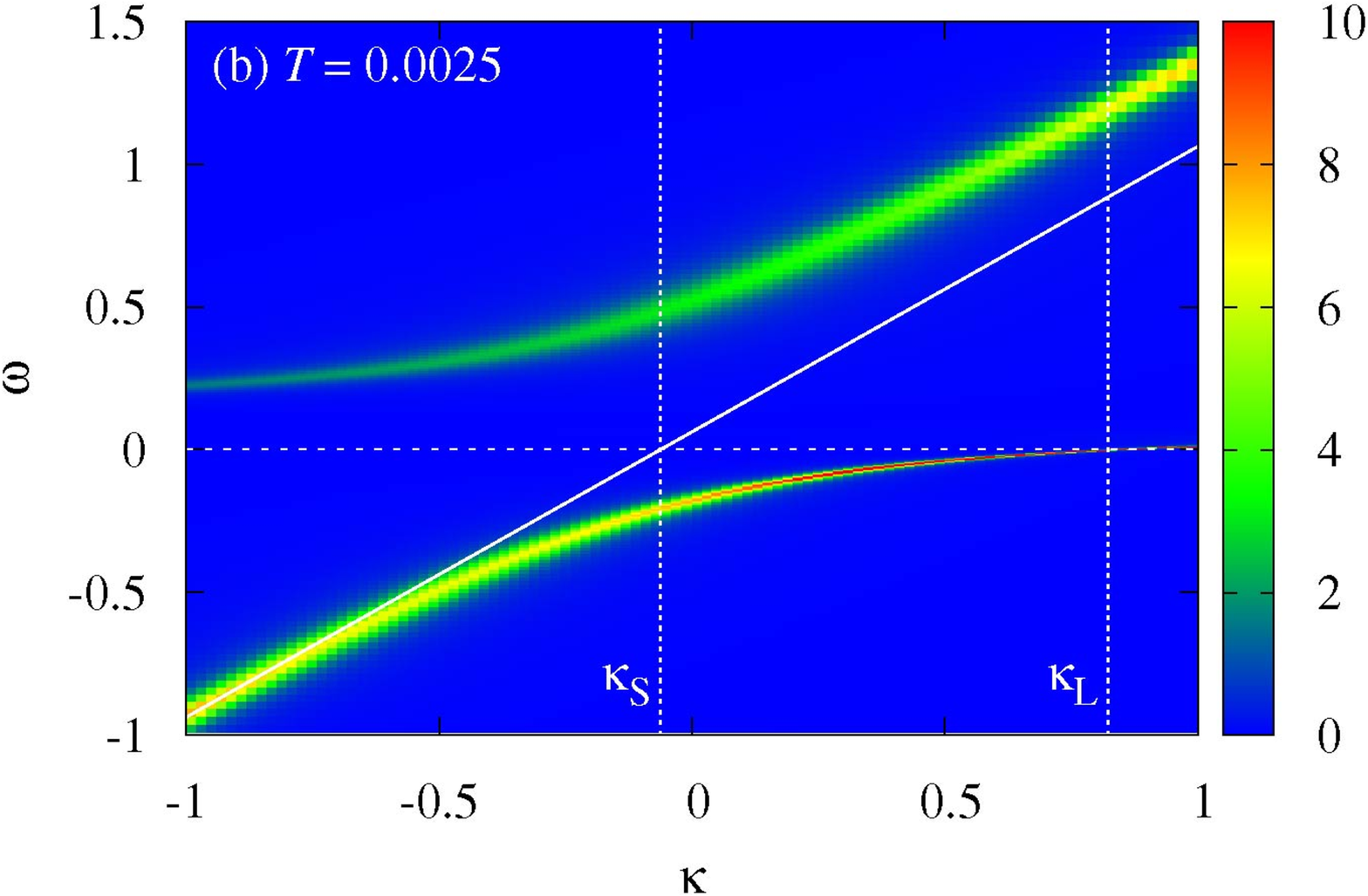}
\caption{The single-particle excitation spectrum 
%$A(\epsilon, \omega)$ 
$A(\kappa, \omega)$ 
for $J=0.3$ and $n_{\rm c}=0.9$ at (a) $T=0.25$ and (b) $T=0.0025$. 
The slanted line represents the non-interacting 
%``dispersion", 
spectrum $\omega=\kappa -\mu$ which is realized with $J=0$.}
\label{fig:spectra}
\end{figure}

{\bf Single-particle spectra.} ---
%Let us examine the single-particle excitation spectra in the high-temperature localized-moment regime and the low-temperature Fermi-liquid state for $J=0.3$. 
%.
In the DMFT, 
${\bm k}$-dependence of the Green function
enters only through $\epsilon_{\bm k}$.
Therefore $G_{\rm c}({\bm k},i\epsilon_n)$ is regarded as a function of $\kappa=\epsilon_{\bm k}$
%. (we call it as ``momentum" 
and is written as
\begin{equation}
G_{\rm c}(\kappa, i\epsilon_n) = [i\epsilon_n -\kappa + \mu - \Sigma_{\rm c}(i\epsilon_n)]^{-1}.
\label{eq:Gc}
\end{equation}
%.omitting the subscript ${\bm k}$.
The single-particle excitation spectrum 
$A(\kappa, \omega) = -\text{Im}G_{\rm c}(\kappa, \omega+i0)/\pi$ is obtained
%we perform the analytic continuation, $i\epsilon_n \rightarrow \omega+i\delta$, in Eq.~(\ref{eq:Gc})
by means of the Pad\'e approximation~\cite{pade}.
%.Note that our 
The validity of the Pad\'e approximation has been confirmed %extensively 
for the impurity Kondo-type models~\cite{Otsuki-CTQMC}.
Our QMC data has the accuracy high enough to obtain reliable real-frequency spectra.

Figure \ref{fig:spectra} shows 
the single-particle excitation spectra for a fixed value of $J=0.3$.
We define two 
%.Fermi momenta, 
energies, $\kappa_{\rm L}$ and $\kappa_{\rm S}$, corresponding to the large and small Fermi surfaces, respectively.
The upper panel 
%in Fig.~\ref{fig:spectra} 
shows $A(\kappa, \omega)$ for $T=0.25$.
The spectrum exhibits a behavior
%.nearly 
of almost non-interacting electrons
%.``dispersion" 
at high energies.
However, the coupling with spin fluctuations 
%.at this temperature already destroy the coherent motion of the conduction electrons yielding 
give rise to a pseudo-gap near the small Fermi surface $\kappa_{\rm S}$.
%.It is obvious 
%We notice
%that the blurry Fermi surface is roughly located at $\kappa_{\rm S}$.
%At lower temperatures, the multiple scattering off the local spins becomes essential. At the low enough temperature, 

%.
The lower panel shows the spectrum at $T=0.0025$, which is much lower than the impurity Kondo temperature defined by 
$
%T_{\rm K}=\sqrt{2J\rho_{\rm c}(0)}\exp[-1/(2J\rho_{\rm c}(0))]\sim 0.1.
T_{\rm K}=\sqrt{g}e^{-1/g} \sim 0.1$ 
with $g=2J\rho_{\rm c}(0)$.
The pseudo-gap looks
%.prominent, 
like a true gap, and the coherent quasi-particles develop near the gap edges.
This structure has a close resemblance to the hybridized band in the Anderson lattice. 
%.(imagine a linear dispersion relation).
In particular, the width of the pseudo-gap is the order of 
the Kondo temperature 
$T_{\rm K}\sim 0.1$.
The quasi-particle intensity becomes
sharper near the Fermi energy, and the quasi-particle band crosses the Fermi energy exactly at $\kappa_{\rm L}$.
Namely, the localized 
%.spin does 
spins do 
contribute to the Fermi volume.
%. via the exchange coupling $J$.
% The width of the pseudo-gap is the order of the Kondo temperature, $T_{\rm K}=\sqrt{2J\rho_{\rm c}(0)}e^{-1/2J\rho_{\rm c}(0)}\sim 0.1$, which essentially differs from the coherence temperature, $T^*\sim0.03$.

{\bf Evolution of the Fermi liquid.} ---
Let us characterize the Green function near the Fermi surface.
We expand the self-energy in powers of $i\epsilon_n$
%$\Sigma_{\rm c}(i\epsilon_n)=\Sigma'_{\rm c}(i\epsilon_n)+i\Sigma''_{\rm c}(i\epsilon_n)$ 
to obtain
\begin{equation}
	G_{\rm c}(\kappa, i\epsilon_n) \simeq \frac{z}{i\epsilon_n - z(\kappa - \mu + 
	\Sigma_{\rm c}(0))}, 
\label{eq:Gc-approx}
\end{equation}
where the renormalization factor $z$ is given by
\begin{equation}
	z = \left( 1 - \left. \frac{\partial 
	{\rm Im} \Sigma_{\rm c}(i\epsilon_n)
	}{\partial \epsilon_n} \right|_{\epsilon_n \rightarrow +0} \right)^{-1}.
\end{equation}
From the QMC data, 
we have confirmed numerically the Fermi-liquid conditions, 
$\partial {\rm Re}\Sigma_{\rm c}(i\epsilon_n)/ \partial \epsilon_n\to0$ and 
${\rm Im}\Sigma_{\rm c}(i\epsilon_n) \to0$ 
in the limit $\epsilon_n\to 0$,
%..
which means $T\to 0$.
As a result, we obtain 
$z\simeq 0.084$ for $J=0.3$.
% From the QMC data of 
% ${\rm Re}\Sigma_{\rm c}(i\epsilon_n)$ 
%...
% at low temperatures and
% for small $|n|$,
%and $\Sigma'_{\rm c}(i\epsilon_1)$,
We extrapolate 
${\rm Re} \Sigma_{\rm c}(0)$ 
from the QMC data of
${\rm Re}\Sigma_{\rm c}(i\epsilon_n)$ 
for small $|n|$, 
assuming a quadratic function centered at $\epsilon_n=0$.
%... 
% for $T\to 0$.

\begin{figure}[tb]
\includegraphics[width=8.5cm]{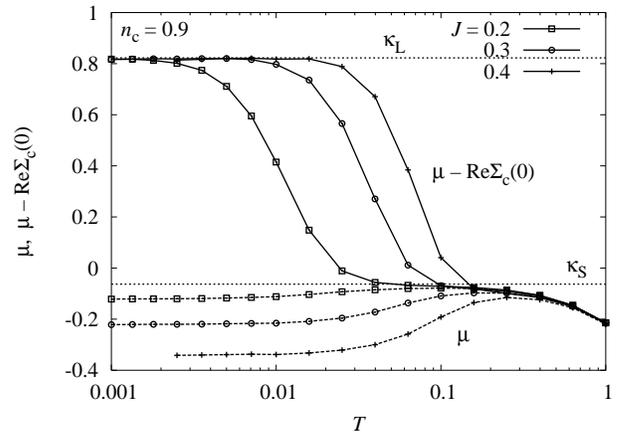}
\caption{
%.The 
Temperature dependences of the chemical potential $\mu$ and its self-energy correction, $\mu-{\rm Re}\Sigma_{\rm c}(0)$, for $n_{\rm c}=0.9$.}
\label{eq:chem_pot}
\end{figure}

We now demonstrate the evolution of the Fermi-liquid state with the large Fermi surface
as a function of temperature. 
The Fermi surface appears at such momentum that satisfies 
$\epsilon_{\bm k} =\mu-\Sigma_{\rm c}(0)$.
Figure \ref{eq:chem_pot} shows temperature dependences of $\mu-{\rm Re}\Sigma_{\rm c}(0)$
and $\mu$ for $J=0.2$, $0.3$ and $0.4$.
There are apparently two regimes in temperature.
At high temperature, the self-energy correction is not important,
and thermal excitations occur around $\mu\sim \kappa_{\rm S}$.
At low temperature, on the other hand, $\mu-{\rm Re}\Sigma_{\rm c}(0)$ deviates significantly from $\mu$, 
and gives rise to the large Fermi surface at $\kappa_{\rm L}$.
From this observation, $\mu-{\rm Re}\Sigma_{\rm c}(0)$ turns out to be a good measure to characterize the crossover between the local-moment regime and the Fermi-liquid state.
%Therefore, it can be said that $\Sigma_{\rm c}(0)$ plays an essential role in turning to the FL state in the KLM. 
%We measure the evolution from the localized moments to the FL states via $\mu-\Sigma_{\rm c}'(0)$. 
%To this end, It is convenient to 
We introduce a dimensionless quantity 
\begin{equation}
%..\xi=(\mu-\Sigma_{\rm c}'(0)-\kappa_{\rm L})/
\xi=(\mu-{\rm Re}\Sigma_{\rm c}(0)-\kappa_{\rm L})/
(\kappa_{\rm S}-\kappa_{\rm L}),
\label{xi}
\end{equation}
which ranges from 0 to 1 for $T \ll D$.
%.With this definition, 
The limit $\xi=0$ 
%and 1 
represents the Fermi-liquid state with the large Fermi surface,
 and the other limit $\xi=1$ represents
 almost non-interacting conduction electrons.
%As has been shown 
%we determine 
We define the crossover temperature $T^*$ 
%.has been defined 
by the condition $\xi=0.5$ as shown in Fig.~\ref{fig:phase_diagram}.
%We define a characteristic temperature $T^*$ as a contour line at
At $J=0.3$, for example, we obtain $T^* \simeq 0.035$,
% which roughly corresponds to 1/3 of $T_{\rm K} \sim 0.1$.
which is smaller than $T_{\rm K} \sim 0.1$.
%At $J=0.3$, we obtain $T^* \simeq 0.035$, which is 
%.the same order of $zD\sim 0.084$, but essentially differs from 
%, which characterizes the pseudo-gap.
%

For the one-dimensional KLM,
the existence of two different energy scales 
%.metallic phase 
has already been discussed in the 
literature~\cite{Shibata-Tsunetsugu}.
The uniform charge susceptibility $\chi_{\rm c}(T)$, for example, takes maximum around $T_{\rm K}$ and decreases with decreasing temperature due to the development of the pseudo-gap.
With a further decrease of temperature, $\chi_{\rm c}(T)$ turns to increase 
%.since particle-hole excitations among the coherent quasi-particle band begin to 
reflecting contribution of coherent excitations, 
and it saturates to a finite value.
For the infinite-dimensional KLM,  our 
%.preliminary 
results of the uniform charge susceptibility, which will be reported elsewhere, shows similar behavior with that in one dimension.
The distinct energy scales without phase transitions are important to analyze experimental data properly.
%%It is believed that the hybridized-band gap in the Anderson lattice is characterized by the coherent temperature $T^*$.
%%In view of connection to the KLM, there may also exist 

%The presence of two distinct energy scales in the magnetically frustrated KLM has also been discussed \cite{burdin}.  The impurity problem has been solved by the saddle-point approximation, which assumes the large degeneracy in the localized degrees of freedom.
We note the difference between the present description in Eq.~(\ref{eq:Gc-approx}) and the ordinary Fermi-liquid state in the Anderson lattice~\cite{Yamada-Anderson-lattice}.
In the Anderson lattice, the localized 
%.
electrons which are practically immobile at high temperatures eventually become itinerant.
%.coherent carriers through the crystal.
Thus, the major component of the quasi-particle is the localized electron.
On the contrary, the local spins in the KLM remain
%.are rigorously 
absolutely immobile although the Fermi surface becomes large.
At high temperatures, the conduction electrons interact incoherently with local spin fluctuations. 
At low temperatures and energies, the conduction electrons manage to form coherent quasi-particles which 
%.are almost scattering free from 
accompany the polarization cloud of spin fluctuations.
%.The resultant ``scattering free" 
These quasi-particles exhibit the heavy-fermion behavior.
In particular,  the origin of the large linear specific heat is the entropy of the local spins as in the Anderson lattice.
The spin entropy changes in the temperature scale of $T_{\rm K}$.

\begin{figure}[tb]
\includegraphics[width=8.5cm]{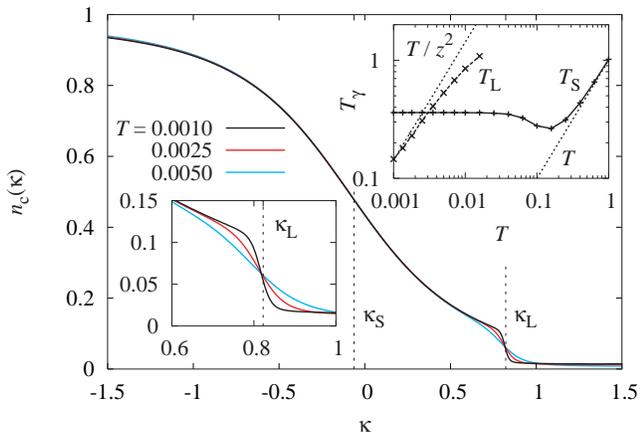}
\caption{
%.The 
Momentum distribution $n_{\rm c}(\kappa )$ for $J=0.3$ and $n_{\rm c}=0.9$. 
The vicinity of the large Fermi surface is enlarged in the left inset.
The right inset shows the temperature dependences of the ``width'' of $n_{\rm c}(\kappa )$ at $\kappa =\kappa_{\rm S}$ and $\kappa_{\rm L}$.}
\label{fig:distrib}
\end{figure}

%.{\bf Momentum distributions.} ---
{\bf Momentum distribution.} ---
%.To elucidate the discontinuity at the Fermi surface,
%make clear the low-energy excitations at $T \ll T^*$,
We now discuss the 
momentum distribution of conduction electrons:
\begin{equation}
\langle c_{{\bm k}\sigma}^\dagger c_{{\bm k}\sigma}\rangle
=T\sum_n G_{\rm c}(\kappa , i\epsilon_n) e^{i\epsilon_n 0_+}
\equiv n_{\rm c}(\kappa ).
\end{equation}
Figure~\ref{fig:distrib} shows 
%.the momentum distribution 
$n_{\rm c}(\kappa )$
in the Fermi-liquid 
%state, 
state, $T \ll T^*$.
The gradient of $n_{\rm c}(\kappa )$ at $\kappa =\kappa_{\rm L}$ becomes steep with decreasing temperature, indicating the existence of 
%.the Fermi-surface 
discontinuity at $T=0$.
The magnitude of the discontinuity is consistent with $z\simeq 0.084$ estimated from $\Sigma_{\rm c}(i\epsilon_n)$.
% As $J$ decreases toward the quantum phase transition at $J=J_c\sim 0.2$, the magnitude of $z$ tends to zero according to
% \begin{equation}
% z \sim \alpha (J-J_c)^\beta
% \end{equation}
% with $\alpha=?$ and $\beta=1?$
On the other hand, $n_{\rm c}(\kappa )$ at $\kappa_{\rm S}$ is almost independent of $T$ in the Fermi-liquid state.
For characterization of the momentum distribution at finite temperature,
we introduce the energy scales ($\gamma={\rm L}, {\rm S}$):
%.of the thermal excitation 
\begin{equation}
T_{\gamma} = \frac{1}{4} \left( \left. -\frac{\partial n_{\rm c}(\kappa )}{\partial \kappa } \right|_{\kappa =\kappa_{\gamma}} \right)^{-1},
\label{eq:TS_TL}
\end{equation}
at $\kappa_{\rm L}$ and $\kappa_{\rm S}$. 
%.This energy scale represents
In the Fermi-liquid state with the approximate Green function, Eq.~(\ref{eq:Gc-approx}), $T_{\rm L}$ should behave as $T/z^2$.
%If the distribution is given by the Fermi distribution function $f(\kappa -\kappa_{\gamma})$,  
% The approximate Green function (\ref{eq:Gc-approx}) with $\mu-\Sigma_{\rm c}(0)=\kappa_{\rm L}$ gives $T_{\rm L}=T/z^2$.
The inset in Fig.~\ref{fig:distrib} shows
the temperature dependences of $T_{\gamma}$.
The linear-$T$ dependence in $T_{\rm L} \simeq T/z^2$ at temperatures lower than $T^*$ ($\simeq 0.035$ for $J=0.3$) verifies the existence of the discontinuity at $T=0$, and ensures the validity of the quasi-particle expansion in Eq.~(\ref{eq:Gc-approx}).
% Namely, this result verifies the temperature-dependent evolution
% %.existence 
% of the large Fermi surface as well as the validity of the low-energy representation in Eq.~(\ref{eq:Gc-approx}).
At temperatures higher than $T_{\rm K}$, on the other hand, $\kappa_{\rm L}$ loses its importance.
Instead,
thermal excitations are populated around $\mu\sim \kappa_{\rm S}$ 
which develops according to $T_{\rm S} \simeq T$.
%However, 
%, which has been derived on the assumption of the Luttinger's relations.

%???The multiple scattering off the localized spins transfers the low-energy spectral weight of the conduction electrons onto the edges of the pseudo-gap.
%As a result, the single-particle excitation spectrum exhibits the hybridized-like quasi-particle band, which crosses the Fermi energy at the large Fermi surface.
%% in spite of the absence of the charge fluctuations in the localized electrons. 
%%As a result, the quasi-particle states appear around the large Fermi surface. 
%The static component of the self-energy, $\Sigma_{\rm c}(0)$, plays a key role to realize the large Fermi surface.
%%The renormalization of the chemical potential plays an important role in the evolution from the localized moments to the FL.
%%The self-energy seems to adjust the Fermi momentum so as to involve the localized spins into the Fermi sea.
%It should be emphasized, however, that a shape of the renormalized Fermi surface is highly non trivial for a realistic energy dispersion in finite dimensions.
%In this case, the ${\bm k}$-dependence of $\Sigma_{\rm c}({\bm k}, 0)$ would be relevant.

We have thus established the Fermi-liquid state of the KLM in infinite dimensions.
On the basis of the Luttinger's theorem, we may further develop a phenomenological Fermi-liquid theory in the vicinity of the large Fermi surface.
Note that the Fermi-liquid state realized in the KLM has no explicit counterpart of non-interacting system.
This situation is identical to the local Fermi-liquid theory by Nozi\`eres in the single-impurity Kondo model~\cite{Nozieres},
where the phase shift of conduction electrons is a key quantity.
Following similar strategy, an extension should be possible starting from the large Fermi surface in the three-dimensional KLM.
Investigation of possible superconductivity is also an challenging issue in extending
the present infinite-dimensional theory for the KLM.

%Landau's Fermi-liquid theory is constructed on the basis of the one-to-one correspondence of the low-energy excitations when an interaction is adiabatically switched on.
%This concept enables us to deduce a shape of the renormalized Fermi surface and low-energy properties.
% However, the Fermi-liquid state realized in the KLM has no explicit counterpart of non-interacting system.
%  and the Fermi surface is drastically different from that in $J=0$.
%How this concept alters %for the Fermi liquid
%in the KLM is an interesting open question.

In summary, we have demonstrated the 
%..
temperature-dependent 
evolution of the Fermi-liquid state with the large Fermi surface.
Our numerical results in the low-temperature limit
confirm the 
%.topological proof 
premise for Luttinger's theorem that relies on 
%.the presence of 
the Fermi-liquid character of low-energy excitations~\cite{Oshikawa}.

We acknowledge valuable discussions with N. Shibata and H. Yokoyama.
One of the authors (J.O.) was supported by Research Fellowships of the Japan Society for the Promotion of Science for Young Scientists.


\begin{thebibliography}{99}
\bibitem{Otsuki-KLM} For Kondo-type models, a calculation of the two-particle Green function is more involved. The detailed description will be published elsewhere.
\end{thebibliography}
\end{document}